# Economic and Policy Uncertainties and Firm Value: The Case of Consumer Durable Goods

Bahram Adrangi
W.E. Nelson Professor of Financial Economics
University of Portland
5000 N. Willamette Blvd.
Portland, Oregon 97203
adrangi@up.edu

Saman Hatamerad
University of Portland
samanhatamerad@yahoo.com

Madhuparna Kolay
University of Portland
5000 N. Willamette Blvd.
Portland, Oregon 97203
kolay@up.edu

Kambiz Raffiee
Foundation Professor of Economics
College of Business University of Nevada, Reno
Reno, Nevada 89557
raffiee@unr.edu

**April 2025**

# Economic and Policy Uncertainties and Firm Value: The Case of Consumer Durable Goods

**Abstract**

The objective of this study is to analyze the response of firm value, represented by the Tobin's Q (Q) for a group of twelve U.S. durable goods producers to uncertainties in the US Economy. The results, based on an estimated panel quantile regressions (PQR) and panel vector autoregressive MIDAS model (PVM), show that Q for these firms reacts negatively to the positive shocks to the current ratio, and debt-to-asset ratio and positively to operating income after depreciation and the quick ratio in most quantiles. The Q of the firms under study reacts negatively to the economic policy uncertainty, risk of recession, and inflationary expectation, but positively to consumer confidence in most quantiles of its distribution. Finally, Granger causality tests confirm that the uncertainty indicators considered in the study are significant predictors of changes in the value of these companies as reflected by Q.

## INTRODUCTION

In the first quarter of 2024, personal consumption expenditures (PCE) made up almost 68% of the U.S. Gross Domestic Product (GDP). PCE is the spending of households and individuals on goods and services, as well as large one-time purchases. This spending is a key driver of the economy and a reflection of economic trends. For example, if consumption grows by 2% steadily, it can contribute almost 1.4 percentage points to GDP growth.

Durable goods are products that are expected to last at least three years, such as appliances, electronics, furniture, automobiles, and sporting equipment. Durable consumer goods play a pivotal role in modern economies. In 2022, US consumers spent 13% of their total expenditures on durable goods, which is $2.2 trillion.

These goods, often characterized by their ability to provide utility over several years, differ significantly from non-durable goods, which are consumed quickly. The production, purchase, and use of durable goods not only reflect consumer confidence and economic health but also contribute significantly to the Gross Domestic Product (GDP) through their multiplier effects.

Durable consumer goods play a vital role in the economy, acting as indicators of economic health, contributors to GDP, and catalysts for the multiplier effect. Their production and consumption are deeply intertwined with economic cycles, influencing and reflecting broader economic conditions. Understanding the significance of durable goods provides valuable insights into economic dynamics and the mechanisms through which consumer behavior impacts overall economic growth.

Durable consumer goods are essential indicators of economic health. Their purchase decisions are often postponed in times of economic uncertainty and accelerated during periods of economic growth. This cyclical behavior makes them a reliable barometer of consumer confidence and overall economic conditions. According to the Bureau of Economic Analysis (BEA), durable goods expenditures account for a significant portion of personal consumption expenditures, which is a primary component of GDP. The demand for durable goods drives production, investment, and employment in related industries, thereby influencing economic cycles.

The production process of durable goods is complex and resource-intensive, involving significant capital investment and labor. This complexity means that fluctuations in the demand for durable goods can have extensive ripple effects across the economy. For example, the automotive industry, one of the largest producers of durable goods, involves a vast network of suppliers and service providers, from steel manufacturers to dealerships and repair shops. Changes in automotive sales can therefore impact multiple sectors, highlighting the interconnectedness of durable goods production with broader economic activity.

Durable consumer goods contribute directly to GDP through personal consumption expenditures, which comprise one of the three main GDP components, alongside business investment and government spending. According to the Federal Reserve, personal consumption expenditures on durable goods have shown a robust increase over the past decades, reflecting their growing importance in the economy. The sale of these goods not only reflects immediate economic activity but also spurs future consumption and investment.

The contribution of durable goods to GDP can be seen in both their production and post-sale phases. The manufacturing of durable goods requires substantial investment in machinery, technology, and human resources, contributing to industrial growth and employment. Once sold, these goods generate additional economic activity through maintenance, repairs, and upgrades, further adding to their economic impact. This sustained contribution underscores the importance of durable goods beyond their initial purchase.

According to Economic Policy Institute, one hundred jobs in the manufacturing of durable goods stimulate seven hundred forty-four jobs in its supply chain and other indirectly related areas.  An additional one million dollars of consumer spending to durable goods results in 1.8 direct and 16.5 indirect jobs in the related sectors of the economy.  Consequently, the demand for and the manufacturing of durables goods triggers a significant multiplier effect

The concept of the multiplier effect is central to understanding the broader economic impact of durable consumer goods. The multiplier effect refers to the phenomenon where an initial injection of spending leads to a greater overall increase in economic activity. In the context of durable goods, this effect is particularly pronounced. For instance, the purchase of a new car does not only benefit the automaker but also stimulates demand for raw materials, parts, and services, creating jobs and income in related sectors. This, in turn, leads to increased spending by those employed in these sectors, further boosting economic activity.

Durable goods also influence the economy through their role in household wealth and financial stability. Ownership of durable goods often constitutes a significant portion of household assets, impacting consumer confidence and spending behavior. During economic downturns, the value of durable goods may decline, affecting household wealth and reducing spending. Conversely, during periods of economic growth, the increased value and acquisition of durable goods can enhance household wealth and stimulate further consumption and investment (Case, Quigley, & Shiller, 2005).

Firms in the durable goods industry are significantly impacted by macroeconomic uncertainties, which are an inherent feature of every economy. Companies in this sector are often required to make critical operational and financial decisions under unpredictable conditions. The relationship between economic uncertainty and firm value is deeply rooted in economic theory.

Early research, such as Sandmo's seminal work in 1971, examined the influence of uncertainty on corporate behavior. Similarly, John Kenneth Galbraith's *The Age of Uncertainty* (1977) underscored the profound impact of uncertainty on financial and economic systems.

Building on this theoretical foundation, scholars like Flacco and Kroetch (1986), Fooladi (1986) and Fooladi and Kayhani (1991), and Adrangi and Raffiee (1999) investigated how firms navigate market risks. These studies shed light on how companies adjust their production processes, pricing strategies, and profit-maximizing efforts in response to economic uncertainties. More recently, innovations by Becker et al. (2012, 2014, 2016) and Golchin and Riahi (2021) in measuring economic and policy uncertainties have enabled researchers to explore the effects of various uncertainty measures on firm behavior across different industries. For instance, Adrangi et al. (2019, 2023a, 2023b and 2024a, 2024b, 2024c, 2025a, 2025b) investigate the association of the uncertainty with the US financial markets, airline industry and crude oil markets. This study builds upon this growing body of literature by extending the analysis to firms in the U.S. durable goods sector, offering insights into how these firms respond to macroeconomic and policy uncertainties.

The remainder of the paper is organized as follows. Section II offers a review of literature pertaining to Tobin's Q. The data and sources of the sample data are discussed in section III. A brief description of the econometric methodology is the subject of section IV. Section V is earmarked for the discussion of the empirical findings and their implications for the market participants. Summary and conclusions comprise the last section.

## REVIEW OF LITERATURE

In this section, we define Tobin's Q and review studies examining the factors that influence a firm's Q, including economic policy and market uncertainties across various industries and economies relevant to this research. The existing literature is largely concentrated on the oil and gas sector, likely due to the availability of data and the relative ease of computing Q for firms in this industry. As a result, we keep our literature review concise.

Tobin's Q is a ratio that measures a firm's market value relative to the replacement cost of its assets or

Q= Market value of a firm's assets/Replacement cost of the same assets

Where,

Market Value of Assets = Market value of equity + Market value of debt

Replacement Cost of Assets = The cost to replace the firm's existing assets at current prices

If Q > 1, The firm is valued more by the market than the cost of replacing its assets, indicating strong growth prospects and potential overvaluation. Coversly, if Q < 1, the firm is valued less than the cost of replacing its assets, suggesting undervaluation or inefficient asset use.  Finally, if Q ≈ 1, the firm's market value is roughly equal to the cost of replacing its assets, implying a fairly valued firm.

Q is widely used by companies and analysts. For instance, a high Q encourages firms to invest in new assets and may also signal potential acquisition targets. Additionally, it reflects market perceptions of a firm's future profitability and operational efficiency.

In summary, Tobin's Q is widely used in corporate finance, macroeconomics, and industrial organization to analyze investment behavior and market efficiency. We provide an overview of key studies across various economic sectors and highlight additional research relevant to this study.

Butt et al. (2023) explore the use and validity of Q as a metric for firm performance, particularly in the context of marketing-related research. The study highlights the widespread use of Tobin's Q among scholars due to its forward-looking nature and comparability across industries. However, authors raise concerns about the metric's reliability, particularly when it is used in research involving intangible assets like marketing, human resources, and R&D.

The study's data comprises 196 publicly traded non-financial firms in Pakistan, spanning various sectors over five years. The data is sourced from financial statements and stock price information, ensuring a robust dataset for empirical analysis. The authors aim to empirically assess whether Q is an appropriate performance metric by comparing it with stock return, a commonly used alternative metric.

Authors conclude that while Tobin's Q remains a reliable metric, but its use in research involving intangible assets should be approached with caution. The authors recommend that marketing scholars consider alternative metrics, such as stock return, or use multiple performance measures to ensure robust and reliable results. This research contributes to the ongoing debate about the appropriate use of performance metrics in marketing and management studies, particularly in emerging market contexts like Pakistan.

Jin and Jorion (2006) investigate the impact of hedging activities on the market value (MV) of U.S. oil and gas companies. The study is grounded in the theories of hedging, which suggest that managing financial risk through hedging should increase firm value by reducing the volatility of earnings and lowering the costs associated with financial distress, taxes, and underinvestment.

Their study sample of 119 U.S. oil and gas producers over the period from 1998 to 2001, provide 330 firm-year observations. The data was collected from the firms' annual 10-K financial reports, with detailed information on the firms' hedging activities, including the use of futures, options, swaps, and fixed-price physical delivery contracts. The study also accounts for the valuation of oil and gas reserves, which are critical assets for these firms.

The study tests whether this reduction in risk translates into higher firm value by analyzing various definitions of Q, which measures the ratio of the firm's market value to the replacement value of its assets. Despite the clear impact of hedging on reducing stock price sensitivity, this risk management strategy does not appear to translate into higher Q ratios for the firms in the sample. This result challenges the prevailing view that hedging necessarily increases firm value and suggests that the benefits of hedging may be more nuanced, possibly dependent on industry-specific factors or the precise nature of the risks being hedged.

Fadul (2005) investigates the relationship between ethical behavior, corporate social responsibility (CSR), and firm value within the oil and gas industry. Their study is based on a sample of 55 companies within the U.S. oil and gas extraction sector. Fadul (2005) uses a series of social indexes—environmental, diversity, community service, and ethical behavior—as proxies for CSR performance. Firm value is measured using Return on Equity (ROE) and an approximate Q ratio, which represents the market valuation of the company relative to its asset replacement cost.

The main findings of this study from multiple regression analysis reveal that ethical behavior is positively and significantly correlated with firm value. Specifically, the diversity index shows a positive correlation with firm value as measured by Q in the equipment and services sector.

Syaifulhaq & Herwany (2020) examine the relationship between a company's capital structure—particularly its debt levels—and its overall value. The study is motivated by the ongoing debate in corporate finance regarding the optimal capital structure that maximizes firm value. The research highlights the importance of understanding how different financial structures impact firm performance, especially in sectors with significant capital needs, such as the oil and gas industry.

The study employs a quantitative research methodology, using statistical tools to analyze the relationship between debt levels (as a percentage of total capital) and firm value, measured through metrics like Q and Return on Equity (ROE).

The important findings of the study reveal a complex relationship between capital structure and firm value. The results suggest that while moderate levels of debt can enhance firm value by taking advantage of tax shields and leveraging profitability, excessive debt tends to have a negative impact on fir value and Q, likely due to the increased risk of financial distress. This finding aligns with the Trade-Off Theory of capital structure, which posits that there is an optimal debt level where the benefits of debt are balanced against the costs.

Yildiz and Kapusuzoglu. (2020) investigate the effects of hedging activities on the performance of oil and gas firms, specifically examining whether these activities add value to the firms. The researchers employ panel regression models to explore the relationship between firm value, proxied by Q, and various measures of hedging activities, such as the use of derivatives.

Their study utilizes data from 76 exploration, production, and integrated oil and gas companies that actively use hedging instruments and report these activities in their financial statements. The data covers a period from 2007 to 2016 and is sourced from the IHS Markit database. The companies included in the analysis follow either IFRS or GAAP accounting standards, ensuring

Their main findings reveal that hedging activities generally have a negative impact on firm value and Q. This result is consistent across different models and is interpreted as an indication that high levels of hedging might signal financial distress or poor management practices to investors. The study also finds that firms with higher debt levels are more likely to engage in extensive hedging, yet this does not necessarily translate into higher firm value. Additionally, the results show that small firms and those more exposed to financial distress are more aggressive in their use of derivatives. The study concludes that while hedging is intended to manage risk, it does not always enhance firm value and its Q. Other scholars including Phan et al. (2014) and Danielsen (2017) also investigate the relationship between hedging activities, the renewable energy and firm valuation and Q in the oil and gas sector.

García-Gómez et al. (2022) analyze the impact of economic policy uncertainty on the performance of US tourism based on a sample of 296 tourism companies from 2000 to 2018 forming a panel sample of 3068 firm-year observations. The empirical findings of their panel regressions estimated by deploying system-generalized method of moments indicate that EPU has a negative impact on return on assets, return on equity, and Tobin's Q. Their findings are consistent for different variable specifications. They also find that firm size and leverage temper the relationship between EPU and firm performance. Their Panel quantile regression estimation show that the association of EPU on US tourism firm performance is asymmetric. Specifically, they show that firms in the 25% quantile of ROA and ROE are less affected by EPU, and that firms in the 75th and higher quantile of Tobin's Q distribution are not affected by the EPU.

Demir, E., & Ersan, O. (2017). This paper examines the effect of Economic Policy Uncertainty (EPU) on cash holding decisions of firms in BRIC countries. By using firm-level data through the 10 years period of 2006–2015, they find that firms prefer to hold more cash when uncertainty increases after controlling for firm level variables with industry and year fixed effects. The results are robust to alternative control variables, EPU calculations and selection of sub-samples. In addition to the country specific EPU levels, global EPU also has a significant positive impact on corporate cash holdings.

The existing body of literature demonstrates a significant scholarly interest in exploring firm valuation, as measured by Tobin's Q, and its relationship with managerial decisions, hedging activities and economic policy or market uncertainties. However, a notable gap persists in the examination of economic and policy uncertainties and their impact on Tobin's Q across various sectors, including the durable goods sector in the United States. The primary objective of the current research is to address this critical gap and contribute to a deeper understanding of these dynamics.

## DATA

The data for this study are derived from Compustat database of WRDS for twelve companies in the durable goods industry for the period spanning from the first quarter of 1980 to the fourth quarter of 2022. The List of these companies is offered in Table 1. The data for some companies in the sample are not available for the entire sample period. In those cases, we extract the available data resulting in an unbalanced panel data.

The quarterly financial variables in the sample data that may impact firm Q are as follows:

Current ratio (CR)

Debt to asset ratio (DA)

Operating income after debt (OIAD)

Quick ratio (QR)

The monthly variables that account for economic policy and market uncertainties that potentially impact the Q are as follows.

Economic policy uncertainty index (EPU)

Recession Risk (RECRISK)

University of Michigan consumer sentiment index as a measure of consumer confidence.

University of Michigan inflation expectation

We present the summary statistics for the variables in the study in Table 2 and provide a brief explanation of the variables below.

Current ratio= Current assets / Current liabilities.  The higher current ratio ensures that a company will be able to meet short term obligations that may be coming due within a year.  It is reasonable that positive shocks to the current ratio, raises the value of a company as well as its Q.

Similar to CR, the rising the quick ratio is expected to have a positive association with Q. The quick ratio measures a company's ability to pay its current liabilities without urgent sale of its inventory or seeking additional financing.

The quick ratio is an indicator of a company's short-term liquidity position and measures a company's ability to meet its short-term obligations with its most liquid assets. QR is defined as

Quick Ratio= Quick Assets/ Current Liabilities
Quick Assets=Cash+CE+MS+NAR,

Where,

CE=Cash equivalents, MS=Marketable securities, and NAR=Net accounts receivable.

Since it indicates the company's ability to rapidly use its assets that can be converted quickly to cash to cover its current liabilities, it is also called the acid test ratio. The quick ratio is considered a more conservative indicator than the current ratio, which includes all current assets as coverage for current liabilities.

OIAD is also showing a positive association with Q. It is intuitively and theoretically plausible that higher OIAD may lead to higher company value and its Q.

The debt-to-asset ratio is a financial metric that compares a company's total debt to its total assets. It is calculated using the formula:

Debt-to-Asset Ratio=Total Debt/Total Assets

This ratio measures the proportion of a company's assets that are financed by debt, providing insight into the company's leverage and financial stability.

DA may boost a company's Q in several ways. A moderate level of debt can help a company leverage its operations. By borrowing, a company can invest in growth opportunities, such as expanding production capacity or entering new markets, potentially leading to higher revenues and profits. This can increase the company's market value and Q.

In addition, interest payments on debt are tax-deductible. This tax shield can reduce a company's taxable income, thereby increasing its after-tax cash flows. Higher after-tax cash flows can enhance the company's value and Q. Finally, taking on debt might signal to the market that the company's management is confident about future cash flows and growth prospects. This can positively influence investor perception and increase the company's market value.

It should be noted that at some level, DA may harm a company's market value and Q. For instance, high levels of debt increase the risk of financial distress and bankruptcy. High debt levels may result in significant interest and principal repayment obligations. These payments can strain a company's cash flows, limiting its ability to invest in growth opportunities and affecting its long-term profitability and value.

Companies with high debt levels have less financial flexibility. In times of economic downturns or unexpected expenses, highly leveraged companies may struggle to secure additional funding, further exacerbating financial difficulties and negatively impacting their value and Q. Excessive debt can lead to higher perceived risk by investors and creditors, resulting in higher interest rates and a higher cost of capital. This can reduce net earnings and diminish the company's overall value and Q.

The impact of the debt-to-asset ratio on a company's value and Q depends on maintaining an optimal level of debt.  In summary, the debt-to-asset ratio is a crucial metric that can either help or harm a company's value and Q depending on how well the company manages its leverage. Moderate and strategically used debt can enhance growth and value, while excessive debt can lead to financial instability and a decline in value.  It appears that the DA has a positive association with the Q of the companies under study.

EPU is provided by Baker et al (2016). It is based on data from ten major US daily newspapers, serves as a tool to measure the impact of media coverage on economic sentiment. Researchers collect articles from digital archives dating back to January 1985, using specific search criteria to identify relevant content related to uncertainty, the economy, and policy.  An article is considered relevant if it includes at least one word from three different sets of terms each. These are "uncertainty" or "uncertain"; "economic" or "economy"; and one of the following policy terms: "Congress", "deficit", "Federal Reserve", "legislation", "regulation", or "White House" (Baker et al. (2016)). The frequency of relevant articles determines the trajectory of the index, which is compiled on a monthly basis (Baker et al. (2012))

The University of Michigan Consumer Confidence Index, officially known as the *University of Michigan Consumer Sentiment Index* (MCSI), is a key indicator of the financial confidence of American households. It is constructed through monthly surveys that gauge consumers' attitudes toward current and future economic conditions.

The *Surveys of Consumers* are conducted by the University of Michigan's Institute for Social Research (ISR).  Each month, approximately 500 telephone interviews are conducted across the U.S. to gather data from a representative sample of U.S. households.

The survey consists of five main questions that measure Current economic conditions and future economic outlook.  Respondents are asked about their current financial situation and their perception of the broader economic climate.

The five questions are as follows:

Personal financial situation now compared to a year ago.

Expectations of personal financial situation one year from now.

Expectations for business conditions over the next 12 months.

Business conditions over the next five years.

Is it a good time to make large purchases, such as major household items?

Responses to the five core questions are used to compute two sub-indices- Current Economic Conditions Index (CCI) based on responses to questions 1 and 5, and Consumer Expectations Index (CEI) based on responses to questions 2, 3, and 4.

The overall *Consumer Sentiment Index (CSI)* is then derived by combining the results of these two sub-indices, where each component is weighted and normalized based on historical data. The baseline value for the index is set to 100 in December 1966.

The index is released twice each month: a preliminary report around the middle of the month and a final report at the end of the month. The final index values provide insights into consumer confidence, reflecting the public's feelings about current and future economic conditions, which can influence spending behavior. The MCSI is widely followed by policymakers, businesses, and investors as a leading indicator of economic trends, particularly consumer spending, which accounts for a significant portion of U.S. GDP.

The University of Michigan's Inflation Expectation is a component of the broader *Surveys of Consumers* conducted by the University of Michigan's Institute for Social Research. It reflects how U.S. consumers expect prices to change over the near and long term, offering valuable insight into future inflation trends from the perspective of the general public.

The inflation expectation data is derived from the *Surveys of Consumers*, the same surveys used for the Consumer Sentiment Index (MCSI).

Each month, about 500 telephone interviews are conducted with U.S. households to gather their views on economic conditions, including expectations for inflation.

Respondents are asked about their expectations for price changes in the short- and long-term, t.e., over the next twelve months and five years, respectively. The questions are open-ended, allowing respondents to provide a percentage figure or qualitative answer about how much they believe prices will change.

Responses are aggregated to produce median inflation expectations for both the 12-month and 5-year horizons. The median (not the average) is used to avoid the influence of outliers, making the measure more robust. These medians represent the expected rate of inflation as perceived by the survey respondents.

The short-term and long-term inflation expectations are reported monthly alongside the broader Consumer Sentiment Index.The data is frequently cited by economists, financial analysts, and policymakers as it provides insight into consumer expectations about inflation, which can influence actual inflation through spending behavior.

Inflation expectations play a critical role in the economy, as they can shape future inflation. If consumers expect prices to rise, they may alter their behavior, such as spending more in the short term or demanding higher wages, which can create a feedback loop that influences actual inflation. The Federal Reserve monitors inflation expectations closely because they impact monetary policy decisions, especially when setting interest rates.

Recession risk is the smoothed U.S. recession probabilities, not Seasonally Adjusted. This series is taken from the Federal Reserve Bank of St. Louis database or FRED. The monthly US recession probability is derived by deploying a dynamic factor model. The model aggregates information from multiple economic indicators to estimate an underlying latent factor that captures the common movements in the data. This factor is assumed to be indicative of the overall economic condition. Key macroeconomic indicators typically include GDP growth, industrial production, employment figures, and other relevant economic variables. The smoothed recession probability is obtained using a Markov switching model, which allows the economy to switch between different states, such as expansion and recession. The model assumes that the economy can transition between these states with certain probabilities, which are estimated from historical data.

The model estimates the probability that the economy is in a recession at any given point in time. This involves calculating the likelihood that the latent economic factor is in a recessionary state based on the observed indicators. The "smoothing" aspect refers to the use of past, present, and future data to provide a more accurate estimate of the recession probability at each point in time.

The smoothed recession probability provides a real-time assessment of the likelihood of a recession, making it a valuable tool for policymakers, economists, and financial market participants. By continuously updating the probability as new data becomes available, it offers a dynamic and timely indication of economic conditions. Businesses and investors can use the smoothed recession probability to inform their risk management and strategic planning decisions. A high probability of recession may prompt firms to adopt more conservative financial practices and prepare for potential economic downturns. Policymakers, such as those at central banks, can use the probability to gauge the health of the economy and decide on appropriate monetary or fiscal measures. For example, a rising recession probability might lead to considerations for stimulus measures to mitigate economic contraction.

Recession probabilities provide a comprehensive measure that incorporates a wide range of economic indicators. The probability information offers a more nuanced and timely assessment compared to single-indicator models. This information is useful for both short-term economic forecasting and long-term strategic planning. The model is not devoid of limitations.

The smoothed recession probability by Chauvet (2008) is a powerful tool for assessing recession risk, leveraging a combination of dynamic factor models and Markov switching techniques to provide timely and informative estimates of economic conditions.

Smoothed recession probabilities for the United States are obtained from a dynamic-factor markov-switching model applied to four monthly coincident variables: non-farm payroll employment, the index of industrial production, real personal income excluding transfer payments, and real manufacturing and trade sales. This model was originally developed in Chauvet (1998,) and further revised by Chauvet and Piger (2008) and Chauvet et al. (2024).

## METHODOLOGY

This section provides a concise overview of the two primary empirical tools utilized in this study: the panel quantile regression and the panel vector autoregressive (PVAR) model. We follow Adrangi et al. (2024a, 2023a, 2023b, and 2025a, 2025b) closely in deploying these methodologies. We also refer readers to Koenker (2004), Canay (2011), Koenker and Ng, (2005)), Sheather (1988) and a kernel bandwidth recommended by Koenker (1987, 2005). Koenker and Machado (1999)  Lamarche, 2010; Galvao, 2011, 2019, and 2020) , and Canova and Ciccarelli (2013). among others,  for more extensive discussion of these methodologies.

**Panel Quantile Regression (PQR)**

Panel data quantile regression (PQR) is a powerful variation of the standard quantile regression methodology (QR) for estimating model coefficients that vary across the conditional distribution of the dependent variable while accounting for both firm-specific and time-specific influences on the dependent variable q.

Panel quantile regression is developed to estimate a conditional quantile regression model with latent individual heterogeneity. Panel data consists of observations on the dependent variable, $y_{it}$, and a $p$-dimensional vector of regressors, $x_{it}$, for subject $i = 1, 2, \ldots, n$ over time $t = 1, 2, \ldots, T$. The work of Lipsitz et al. (1997) and contributions of Koenker (2004) have advanced the QR methodology and inspired new developments in recent decades.

This approach is advantageous over traditional regressions that focus on the conditional mean, as it can capture heterogeneity more effectively. In recent decades, advancements in panel data methodologies have introduced several algorithms, enabling more robust and informative empirical analyses that address variable heterogeneity and factor structure. PQR captures the heterogeneity in the effects of explanatory variables across the distribution of the outcome. This is important because the relationship between variables is rarely uniform across the entire population. For example, the return on education might be higher for people at the top end of the wage distribution and lower for those at the bottom. By only analyzing the mean effect (as in OLS), these varying impacts would be missed.

PQR offers a robust and flexible tool for understanding how explanatory variables influence different parts of the outcome distribution. By examining the effects at various quantiles, one uncovers heterogeneous impacts, gain insights into inequality, manage risk more effectively, and develop targeted strategies. Whether analyzing income inequality, housing prices, or financial returns, quantile regression provides a more comprehensive picture than mean-based methods, making it an essential technique for understanding complex, real-world data.

While it shares the same underlying methodology as QR and offers similar advantages, PQR is tailored to account for the structure and characteristics of panel data. Koenker and Bassett (1978) and Koenke et al. (1987) in their seminal work "Regression Quantiles", created a literature that offers informative alternative to regressions that estimate the conditional mean of a variable as a function of explanatory variables. QR is a technique that estimates covariate effects at different quantiles of the conditional distribution of the dependent variable. This approach provides more nuanced map of the association between a dependent variable and its explanatory variables than the variations of the least squares method. Since its introduction, scholars like Portnoy (1991), Powell(1986), Koenker (2004), He and Zhu (2003), Angrist, et al. (2006), Golchin, Rekabdar. (2024), among others generalized quantile regression, inspiring the deployment of the methodology in many areas of inquiry including finance and economics. A natural extension of the standard QR methodology is its application to panel data.

The key insight here is that the relationship between the independent variables and the dependent variable may differ across different points of the outcome distribution.

Panel quantile regression allows for a more nuanced understanding of inequality by revealing how explanatory variables affect different segments of a population. Policymakers and researchers can use this information to design more targeted interventions. For example, knowing how economic shocks disproportionately affect lower the q distribution can help create policies that protect the most vulnerable groups.

Since PQR estimates are less sensitive to outliers than OLS, they provide more reliable estimates in datasets where extreme values might unduly influence the results. This is especially noteworthy because panel data in general and the one use in this study consists of firms that exhibit diverse q values.  that For example, in this data set, a few extremely high or low q values can distort the mean. PQR helps overcome this limitation by providing a broader picture across different parts of the distribution.

Following the approach of Adrangi et al. (2023a an b), we employ panel quantile regressions (PQR). PQR models, known for their nonlinearity (as seen in Galvao(2011) and Galvao et al., 2019 and 2020), exhibit robustness in the presence of extreme events and asymmetric dependence, where linearity assumptions may be inadequate (as observed in Geraci, 2019; Yu et al., 2003). Additionally, they outperform Ordinary Least Squares (OLS) estimates by allowing coefficient estimates to vary with the distribution of the dependent variable, providing an accurate representation of the relationship between explanatory variables and the dependent variable. The following provides a concise overview of Quantile Regression (QR) which is the basis for the PQR.

Given a random variable $Y(q)$ with probability distribution function

$$F(y) = \text{Prob}(Y \leq y)$$

so that for $0 < \tau < 1$ the $\tau$th quantile of $Y$ is defined as the minimum $y$ satisfying $F(y) \geq \tau$:

$$Q(\tau) = \inf\{y : F(y) \geq \tau\}.$$

Then the empirical distribution function is

$$F_n(y) = \sum_s l(Y_i \leq y),$$

in which $l(.)$ is a binary function that takes the value 1 if $Y_i \leq y$ is true and 0 otherwise. The resulting empirical quantile may be expressed as an optimization problem as

$$Q_n(\tau) = \arg\min\left\{\sum_i \rho_\tau(Y_i - \omega)\right\},$$

in which $\rho_\tau(w) = w(\tau - 1(w < 0))$ asymmetrically assigns weights to positive and negative values in the estimation process.

We assumed a linear specification for the conditional quantile of the dependent variable Q given values for the vector of explanatory variables $X$ such that

$$Q(\tau | X_i, \beta(\tau)) = X_i' \beta(\tau). \tag{1}$$

In equation (1), $\beta(\tau)$ is the vector of coefficients associated with the $\tau$ th quantile.
Then, the conditional quantile regression estimator is as shown by Adrangi et al. (2023a)

$$\beta_n(\tau) = \arg\min_{\beta(\tau)} \left\{ \sum_i \rho_\tau(Y_i - X_i'\beta(\tau)) \right\}.$$

The quantile regression estimator is derived as the solution to a linear programming problem. We apply an adapted form of Koenker and D'Orey's (1987) iteration of the Barrodale and Roberts' (1973) simplex algorithm.

In the last decade, there has been a growing emphasis on introducing more adaptable forms of latent heterogeneity in panel data models. This trend follows significant advancements in the linear panel data literature, which have enhanced the estimation of models with interactive effects and factor structures (Pesaran, 2006; Bai, 2009; Moon and Weidner, 2015, 2017). In the context of panels with a large number of time periods T, panel quantile estimation is performed through a specific optimization process as

$$\arg\min_{\theta, f, \beta \in G \times F \times B} E(\lambda_\tau(y - \mathbf{x}'\beta - \varphi'\mathbf{f})),$$

Where $\varphi$ is a $r$-dimensional vector of individual factor loadings, $r$ is the number of latent factors and $\mathbf{f}$ is a vector of unobserved time-varying factors,. In this formulation the individual effects enter the model such that $\varphi'\mathbf{f} = \varphi_1 + \sum_r \varphi_k f_k$ if $f_1$ is normalized to one.

In this paper we opt for this approach to incorporating subject-specific heterogeneity $\varphi$ and avoid the problem of additive separable unobserved heterogeneity assuming that $\mathbf{f}$ affects the dependent variable.

Consider the panel quantile regression with fixed effects given by equation (2).

$$Qy_{it}(\tau \mid X_{it}, \varphi_i) = x_{it}'\beta(\tau) + \varphi(\tau)_i, \text{ where } i = 1,\ldots,n; t = 1,\ldots,T. \qquad (2)$$

Estimating the quantile regression in (2) presents a problem of incidental paramters as there are incidental parameters that cannot be eliminated.

Kato et al. (2012) studied the properties of quantile regression in a model where the fixed effects are captured by dummy variables. The estimator is consistent and asymptotically normal when both n → ∞ and T → ∞ with $n^2[\ln(n)]^3/T \to 0$. However, this approach leads to explosive conditions as in practice n > T. For instance if T = 30, n = 200, then $n^2[\ln(n)]^3$ = 133,862. An alternative is to deploy the method of moments estimator to equation (2).

$y_{it}$ may be written as

$$y_{it} = \alpha_i + x_{it}'\beta + (\gamma_i + x_{it}'\delta)u_{it}, \text{ where}$$

$$\varphi(\tau)_i = \alpha + \gamma_i Q_u(\tau)$$
$$\text{and}$$
$$\beta(\tau) = \beta + \delta Q_u(\tau).$$

Assuming that $u_{it}$ and $x_{it}$ are independent and that

$$\Pr ob((\gamma_i + x_{it}'\delta) > 0) = 1,$$

we can estimate the model parameters using two fixed effects regressions and deriving a univariate quantile. We deploy this methodology to estimate the parameters of equation (4). To confirm and examine the robustness of our findings, we also estimate panel VAR-MIDAS models. The objective of this exercise is to derive impulse responses of the Q to various shocks to the model variables.

**Panel VAR- MIDAS**

The second tool of our methodology is the panel vector autoregressive MIDAS model (PVM). PVM models are an extension of the standard VAR models that are useful in the presence of panel data and mixed interval data sampling. Our data set is comprised of a mix of quarterly and monthly sample observations that can be analyzed by deploying PVM without data conversions to the same frequency. Similar to VARs, PVM models are particularly adept at modeling systems of interrelated variables. Since standard VARs are extensively described and utilized in the literature, we only offer a summary of this methodology. The reduced form PVM that we deploy avoids the complex structure of modeling the interactions of variables by treating all model variables as endogenous. This approach is superior to others that require pre-determined role as exogenous and endogenous variables because the interconnection among economic and financial variables is not completely cut and dry.

A VAR model of order p may be expresses as equation (1)

$$y_t = A_1 y_{t-1} + A_2 y_{t-2} + \ldots + \varepsilon_t ,\tag{3}$$

where,

$y_t = (y_{1t}, y_{2t} \ldots y_{jt})'$ is a vector of Jx1 model endogenous variables,

$A_1 A_2 \ldots A_p$ is the matrix of model coefficients that will be estimated.

$\varepsilon_t = (\varepsilon_{1t}, \varepsilon_{2t} \ldots \varepsilon_{jt})$ is a j by1 vector of white noise model innovations with the following matrix of variances and characteristics.

$$E(\varepsilon_t) = 0, E(\varepsilon_t \varepsilon_t') = \Sigma_\varepsilon$$

$E(\varepsilon_t \varepsilon_s') = 0$ for $t \neq s$.

The basic idea behind PVM models is to use lagged values of high-frequency data to explain the current or future values of low-frequency data. This is achieved by estimating a dynamic relationship between the variables using a combination of high-frequency and low-frequency data.

We offer a brief explanation of a simple PVM with m high frequency periods per low frequency observation. For instance, for each monthly observation on a variable, there may be twenty daily observations on another set of variables. Its application to the panel data adds another layer of complexity. However, the only difference between the PVM and VAR-MIDAS models is that the low frequency data consists of an unbalanced panel.

To demonstrate a PVM model, let us assume that a VAR consists of $n_{Lt}$ variables observed at low frequency and $n_{Ht}$, at high frequency. Forming a VAR from the high and low frequency variables in a VAR produces the system of equation (2)

$$\begin{bmatrix} Y_{H.t_t,1} \\ Y_{H.t_t,2} \\ \ldots \\ Y_{H.t_t,n} \\ Y_H, t_L \end{bmatrix} = \begin{bmatrix} \Gamma_j^{1,1} & \Gamma_j^{1,2} & \ldots & \Gamma_j^{1,n} & \Gamma_j^{1,n+1} \\ \Gamma_j^{2,1} & \ldots & \ldots & \ldots & \Gamma_j^{2,n+1} \\ \ldots & & & \ldots & \\ \Gamma_j^{n,1} & \ldots & \ldots & \Gamma_j^{n,n} & \Gamma_j^{n,n+1} \\ \Gamma_j^{n+1,1} & \ldots & \ldots & \Gamma_j^{n+1,n} & \Gamma_j^{n+1,n+1} \end{bmatrix} \begin{bmatrix} Y_{H.t_t-j,1} \\ Y_{H.t_t-j,2} \\ \ldots \\ Y_{H.t_t-j,n} \\ Y_{L.t_t-j} \end{bmatrix} + \begin{bmatrix} E_{H.t_L,1} \\ E_{H.t_L,2} \\ \ldots \\ E_{H.t_L,n} \\ E_H, t_L \end{bmatrix} ,\tag{4}$$

where matrix $\Gamma_j^{a,b}$ is $K_H*K_H$, $\Gamma_j^{n+1,b}$ $K_L*K_H$, $\Gamma_j^{a,n+1}$ $K_H*K_L$ for all j, a,b= 1,2,.. n, and $\Gamma_j^{n+1,n+1}$ is $K_L*K_L$.

In this model, $Y_{Ht}$ and $Y_{Lt}$ represent the high and low-frequency variables observed at time t. The lagged values of Y, denoted as Y(t-j), capture the autoregressive component of the low-frequency variable. The coefficient matrices $\Gamma_j^{\alpha,\beta}$ represent the parameters to be estimated, and $E_{H,t_L}$ and $E_{L,t_L}$ are the high and low frequency white noise error term vectors.

Ghysels et al. (2004 and 2016) offers the Bayesian PVM estimation. This methodology requires specifying prior distributions for the model coefficients and the covariance matrix of the white noise residuals.

Combining the specified priors with VAR likelihood generates the conditional posteriors of the model coefficients as well as the innovations covariance matrix as

$(\beta|y,\Sigma) \sim N(\bar{\beta},\bar{\Omega})$ and

$(\Sigma|y,\beta) \sim IW(\bar{Q},\bar{\varpi})$,

where,

$\bar{\Omega} = (\underline{\Omega}^{-1} + (\hat{\Sigma}^{-1} \otimes (X^{"}X^{'})^{-1}))^{-1}$

$\bar{\beta} = \bar{\Omega}(\underline{\Omega}^{-1}\underline{\beta} + (\hat{\Sigma}^{-1} \otimes (X^{"}))y)$

$\bar{Q} = \underline{Q} + E^{'}E$

$\bar{\varpi} = \underline{\varpi} + T$

The model parameters' prior distribution adheres to Litterman's (1986) approach, where the prior means are set to zero except for the first lag terms. Ghysels et, al (2016) introduces modifications to Litterman's assumptions by assigning a distinct hyper-parameter to each own lag term based on its observation frequency. Additionally, the high-frequency series variables are assumed to follow an AR (1) process, leading to exponential parameters in the low-frequency domain. The prior variance of the model coefficients also incorporates cross-frequency variances.

PVM models offer several advantages in econometric modeling. They allow for the incorporation of high-frequency data, which captures more detailed information and short-term dynamics, into the analysis of low-frequency variables. This enables researchers to better understand the relationships and transmission mechanisms between economic variables at different time scales.

In summary, PVM models are a valuable tool in econometric modeling, particularly when dealing with mixed frequency data. They allow for the integration of variables observed at different frequencies and provide insights into the dynamic relationships between them. By capturing the short-term and long-term dynamics of economic variables, VAR-MIDAS models enhance our understanding of complex economic systems.

To ensure that the model variables are stationary, we deploy two standard tests, the Augmented Dickey-Fuller and Phillips-Perron and report their results in Table 2.

In the following, we discuss the empirical findings of this paper and propose managerial strategies that may effectively safeguard firm value in both the short and long term.

## EMPIRICAL FINDINGS AND THEIR MANAGERIAL IMPLICATIONS

Prior to estimating the PQR we present the Levin, Lin, and Chu t* test (LLC) results in Table 2 to test the stationarity of the financial variables in the panel sample data. This test indicates that DA, OIAD, and Q are non-stationary and have to be transformed by first differencing to become stationary. Thus, these variables will enter the PQR and other statistical models in this study in first differences. ADF and PP tests of stationarity the monthly uncertainty series show that uncertainty variables are stationary.

**Findings of the Panel Quantile Regressions**

The quantile regression to be estimated is equation (5)

Q= f(cr, da, oiad, qr, epu, rec_risk, infexp, consconf). (5)

Table 3 report the results of Markov Chain Mote Carlo (MCMC) the PQR estimation of the equation (5). Three lags of the uncertainty variable are included to account for the dynamic relation between these variables and Q of the firms under study. Furthermore, this approach provides the necessary information to perform the Granger Causality tests.

Examining Table 3 shows that variables qr, da, and oiad show a positive association with Q, while the association of cr with Q is negative

Turning to the association of the uncertainty variables with Q we see that with the exception of the consumer confidence, the remaining lagged uncertainty variables are negatively associated with Q in all quantiles.

The estimation results of the PQR and understanding the tails of a distribution (low and high quantiles) is critical for managing risk. For example, knowing how economic uncertainties affect worst-performing or best-performing firms in the industry would help the management of these firms to perform risk analysis, improving decision-making by highlighting potential outcomes under different conditions.

Table 3 shows that the relationship between the uncertainty variables and the Q of the firms under study is not constant across the distribution. The estimated PQR coefficients are unanimously statistically significant in all quantiles of the distribution of the Q. At the lower quantiles, results show how uncertainty variables impact the lower end of the Q distribution. For instance, it reveals how uncertainty affects the firms with low market value relative to their replacement cost.

The estimation results confirm that the uncertainty represented by the EPU, recessionary risks, unexpected inflation, and consumer confidence are strongly associated with Q especially in the higher quartiles. This is especially useful in analyzing high-performing firms, or the upper extremes of the Q distribution. It identifies how uncertainty variables impact high-performance firms with high market value, which is critical for understanding the resilience of these firms to market uncertainties. These findings are informative as firms with higher market value are more susceptible to uncertainties in the marketplace.

The 50th percentile, or median quantile, offers a robust measure of central tendency that is less sensitive to outliers than the mean. The results here indicate the "typical" relationship, making it useful when extreme values may skew the average but not the median. Median regression provides insights similar to OLS but is more resistant to the effects of outliers in the data. It can also identify inequalities or vulnerabilities that are masked when focusing only on the mean, such as how economic shocks might disproportionately affect low Q or underperforming firms and high Q or high-performance firms.

Table 3 shows that at both $50^{th}$ and the $75^{th}$ quantiles the uncertainty variables are consistently associated with the Q and exhibit mostly the expected signs, i.e., as risk shown by each uncertainty variable and recessionary risk rises, the Q of the high-performing firms falls. In the case of the consumer confidence, this association is mostly positive and statistically significant confirming that the Q of the firms in the $50^{th}$ and $75^{th}$ quantile of the Q distribution moves in the same direction as the consumer confidence. The findings of the association between firms' q and the uncertainty and consumer confidence are slightly weaker for the firms in the low q quantile. This could be a sign that the weak-performing firms are not as responsive to markets forces and uncertainties and the high-performing ones.

Uncertainty regarding government policies, such as changes in tax regulations, trade policies, or industry-specific regulations, can create challenges for companies in planning and decision-making. Investors may respond to policy uncertainties by discounting the future cash flows of a company more heavily, resulting in a lower valuation and Q.

Economic policy uncertainties can erode consumer and investor confidence. Reduced confidence may result in decreased spending, investment, and overall economic activity. This lack of confidence can lead to a sell-off in the stock market, affecting equity values and, consequently, the Q of companies. The demand for petroleum products such as gasoline and natural gas is inelastic in the short-run, however, given time, industries and consumers adapt by changing their behavior such that the long-run demand urns elastic and sensitive to economic downturns and policy uncertainties. Investors may reevaluate their expectations for companies in this industry, leading to lower Q.

Secondly, economic uncertainty can disrupt demand and supply chains, affecting companies that and their revenue streams or source materials for their production processes. Additionally, economic policy uncertainty may impact currency exchange rates, interest rates, and inflation levels, all of which can have significant implications for businesses involved in production and supply of durable goods. Fluctuations in exchange rates can affect the competitiveness of exports and imports, while changes in interest rates and inflation can impact borrowing costs and consumer purchasing power, respectively, further affecting companies' financial performance and enterprise value.

Moreover, uncertainty about the future economic landscape may lead to cautious consumer and investor behavior, resulting in reduced spending, delayed investment decisions, and lower revenue generation for companies across various sectors. This decrease in economic activity can have a cascading effect on corporate earnings and ultimately impact the valuation of companies in the form of lower stock prices or reduced acquisition premiums. A negative pattern is also observed in the response of the Q to shocks to the recessionary risks.

During a recession, consumer spending tends to decline, and businesses may experience lower demand for their products and services. This can lead to a decrease in the company's revenue and profitability. Investors often value companies based on their future expected cash flows, and a recession can raise concerns about the firm's ability to maintain or grow its earnings, negatively affecting its enterprise value.

Economic downturns normally increase the risk of default on debt obligations for companies, especially those with high levels of debt. As recessionary pressures mount, companies may struggle to meet their interest and principal payments. This heightened default risk can result in a decrease in the company's credit rating, leading to higher borrowing costs and a potential decline in enterprise value.

Fears of recession and economic policy uncertainties can contribute to increased market volatility. Investors may become more risk-averse, leading to higher market discount rates. Elevated volatility can negatively impact stock prices and valuation multiples, affecting the enterprise value of companies across various sectors.

During a recession, central banks may implement monetary policies such as lowering interest rates to stimulate economic activity. While this can reduce borrowing costs for companies, it may also lead to concerns about the overall economic health. Investors may interpret interest rate cuts as a response to economic challenges, contributing to negative sentiment and potentially impacting company Q values.

Inflationary expectations can significantly influence the demand and supply of consumer durables, which are goods intended to last for several years, such as automobiles, appliances, and furniture. These effects can be both positive and negative, depending on the economic context and consumer behavior.

When consumers expect prices to rise in the future, they may accelerate their purchases of durable goods to avoid higher costs later. This anticipation of inflation can lead to a surge in demand as households seek to lock in current prices. For example, if consumers expect inflation to drive up the cost of cars or home appliances, they might decide to buy these items sooner rather than delaying their purchases. This increased demand can stimulate economic activity and lead to higher sales volumes for manufacturers and retailers of durable goods.

However, inflationary expectations can also have a dampening effect on demand for consumer durables. If consumers believe that their purchasing power will decline due to rising inflation, they might become more cautious with their spending. Concerns about future income and the higher cost of financing durable goods through loans or credit can lead to reduced consumption. For instance, higher expected inflation may increase interest rates, making borrowing more expensive. This can deter consumers from taking out loans for big-ticket items, thereby decreasing demand.

On the supply side, manufacturers might increase production in response to rising demand driven by inflationary expectations. Anticipating higher future costs for materials and labor, producers may ramp up their output to take advantage of current lower production costs. This proactive approach can help businesses maintain profit margins and market share. Additionally, the prospect of increased sales can encourage investment in new technologies and expansion of production capacities, boosting overall industry growth.

Conversely, inflationary expectations can create challenges for the supply side of consumer durables. Rising costs for raw materials, labor, and energy can squeeze profit margins, making it difficult for manufacturers to maintain current production levels without raising prices. Supply chain disruptions and increased volatility in input prices can also lead to production delays and higher operational risks. Furthermore, if businesses anticipate that inflation will erode consumer purchasing power significantly, they might become hesitant to invest in new capacity, fearing insufficient demand in the long term.

The overall impact of inflationary expectations on the demand and supply of consumer durables depends on the balance between these positive and negative effects. In an environment of moderate inflation, the stimulative effects on demand might outweigh the negatives, leading to increased economic activity. However, in cases of high or runaway inflation, the negative consequences, such as reduced consumer purchasing power and higher production costs, could dominate, resulting in decreased demand and supply constraints.

In sum, inflationary expectations play a crucial role in shaping the market dynamics of consumer durables. While they can stimulate demand and prompt manufacturers to increase production in the short term, they can also lead to cautious consumer behavior and supply chain challenges, highlighting the complex interplay between inflation and economic activity.

Consumer confidence, which reflects the overall sentiment of households regarding their financial situation and the economy's future, plays a critical role in shaping the supply and demand dynamics for consumer durable goods. The association between consumer confidence and the market for durable goods can be observed through both positive and negative lenses, impacting economic activity in significant ways.

High consumer confidence typically translates into increased demand for consumer durables. When consumers feel optimistic about their job security, income stability, and the broader economic outlook, they are more likely to make significant purchases, such as automobiles, appliances, and furniture. This optimism encourages households to spend on durable goods, confident in their ability to afford these items and manage any associated debt or financing. For instance, a surge in consumer confidence might lead to higher car sales or increased home improvement projects, driving up demand for related durable goods.

Conversely, low consumer confidence can have a dampening effect on the demand for consumer durables. When consumers are pessimistic about their economic prospects, they tend to cut back on discretionary spending and delay or forgo large purchases. This cautious approach is often driven by fears of unemployment, declining income, or economic recession. For example, during periods of low consumer confidence, households might postpone buying a new car or investing in major home appliances, leading to a decrease in demand for these goods.

On the supply side, strong consumer confidence can incentivize manufacturers to ramp up production of durable goods. Anticipating robust demand, companies may increase their output to meet expected sales, invest in new technologies, and expand their production capacities. This proactive stance not only helps manufacturers capitalize on the favorable market conditions but also stimulates economic growth through job creation and higher business investments. For example, a car manufacturer might introduce new models or increase production volumes in response to sustained high consumer confidence.

However, low consumer confidence can negatively impact the supply of consumer durables. When manufacturers perceive a drop in consumer sentiment, they may scale back production to avoid excess inventory and financial losses. This conservative approach can lead to reduced investments in new technologies or production facilities, potentially stifling innovation and long-term growth in the industry. Additionally, decreased production can result in job cuts and lower economic activity in regions heavily reliant on manufacturing durable goods.

The overall impact of consumer confidence on the supply and demand for consumer durable goods depends on the interplay of these positive and negative factors. High consumer confidence generally fosters a virtuous cycle of increased demand, higher production, and economic growth. In contrast, low consumer confidence can lead to a downward spiral of reduced spending, lower production, and economic contraction.

Policymakers and business leaders closely monitor consumer confidence indicators to gauge economic health and make informed decisions that can stabilize or stimulate the market for durable goods. The association of the consumer confidence with Q of the firms in the sample is expected to be positive. However, in all quantiles reported in Table 3, the coefficient of the lagged consumer confidence index is unexpectedly negative and significant. This anomaly does not justify deleting the consumer confidence variable from equation (1).

**Impulse Responses from PVM Model**

Prior to estimating equation (1) we examine AIC and SC to confirm the optimum lag length of the explanatory variables for PVM and the Granger causality tests. These criteria suggest that the optimum lag length in the model is 2 or 3. We include the lag length of three for uncertainty variables to ensure that the model is not mis-specified.

The PVM estimation produces impulse responses of the Q to the financial and risk variables under study. The variance inflation factors show that EPU, recessionary risk and consumer confidence are weakly correlated causing singularity in the matrix of the variables and preventing the estimation of the PVM. To remedy this problem, we only included the consumer confidence index and the inflation expectations in the final estimation of the PVM.

Firms in the durable consumer goods sector are highly sensitive to macroeconomic uncertainties and risks. Economic policy uncertainty and inflationary pressures create unpredictable operating environments, while recession risks and shifts in consumer confidence directly impact demand for their products. By understanding these factors, firms can better anticipate challenges, adjust strategies, and safeguard their profitability during uncertain economic times.

Figure 1 presents the impulse responses of Q to positive shocks to the PVM variables. The impulse response of the Q to a positive shock to the inflation expectations is downward sloping. Conversely, one standard deviation shock to the consumer confidence results in positive reaction in the Q. These observations are plausible and corroborate the estimation results of the PQR.

Analyzing the IR of q to one standard deviation shock to financial variables shows that the one standard deviation positive shock to cr, qr, and oiad triggers positive reactions in q. These findings bolster the findings of the PQR analysis. However, the positive shocks to the DA show a negative reaction in q, which is also consistent with PQR estimation findings.

**Managerial Implications of the Findings**

The implication of these findings is that the management of the firms with high Q would need to be better prepared for the risks that could impact the value of their firms. While some of these risks like EPU may be harder to manage, firms can hedge against inflationary expectations and falling consumer confidence by actively engaging in options and futures markets. For instance, in the face of expected rise in inflation, options or futures contacts would protect producers of durable good against rising input prices. Similarly, with initial signs of potential future changes in consumer confidence, firms that produce durable goods may take steps to survey their retail market and possibly slow down production to avoid inventory buildup costs and forced labor layoffs.

We show that firms in durables good industry in the US face significant risks from inflation, recession, and changes in consumer confidence, but they can mitigate these risks through effective hedging strategies. By employing commodity hedging, managing costs, diversifying revenue streams, building liquidity, and using marketing tactics to sustain demand, companies can protect their bottom lines and remain resilient during periods of economic uncertainty. These proactive strategies enable firms to weather volatile conditions and continue delivering value to stakeholders.

Inflationary risks arise when the purchasing power of money declines, leading to rising input costs and price volatility. Producers of durable goods rely on raw materials (e.g., metals, oil, agricultural products). By using commodity futures contracts, firms can lock in prices for key inputs in advance. This practice allows them to avoid sudden cost increases and better forecast their expenses.

To protect margins, firms may include inflation-adjustment clauses in contracts with suppliers and customers. These clauses allow firms to raise prices in response to rising costs, keeping profitability intact. For instance, firms may include escalation clauses in contracts that allow them to adjust prices based on changes in material costs like lumber, metals, and others.

Firms can also reduce exposure to inflation by diversifying their supplier base or substituting more cost-effective or stable inputs. This minimizes the risk of being locked into high-cost suppliers during inflationary periods. For instance, may source ingredients from multiple suppliers across regions, ensuring a steady supply at competitive prices.

In some instances, firms can pass rising costs to consumers by increasing product prices. While this strategy may depend on the elasticity of demand, firms with strong brand loyalty or market power can raise prices without significantly affecting demand. For instance, luxury goods firms or firms with highly differentiated products may raise prices during inflationary periods, knowing that demand will remain relatively inelastic.

Firms in durable goods industry are cyclical industries are particularly vulnerable to recessions. Recessions are characterized by a decline in economic activity, reduced consumer spending, and tighter credit conditions. Several Strategies to hedge against recession risks may be deployed.

Firms should focus on maintaining strong liquidity by building cash reserves or obtaining lines of credit to weather periods of reduced demand. Adequate liquidity ensures firms can meet operational costs, service debt, and even invest in opportunities that arise during downturns.

Many large firms maintain rainy-day funds, allowing them to continue operations and meet payroll even when sales decline.

A firm can protect itself from the effects of a recession by maintaining a flexible cost structure, particularly through the use of variable costs. By outsourcing, hiring temporary workers, contract labor, or using pay-for-performance models, firms can scale down quickly when demand drops.

During recessions, borrowing costs can rise, and access to capital can become limited. Firms can manage recession risks by reducing high levels of debt or by hedging against rising interest rates through interest rate swaps or by securing fixed-rate loans. A firm that expects recessionary pressure on credit markets may negotiate long-term fixed-rate loans or use interest rate derivatives to limit exposure to variable interest rate hikes.

Consumer confidence reflects the degree of optimism or pessimism that consumers feel about their financial situation and the overall economy. Fluctuations in consumer confidence can directly affect demand for products, especially in discretionary durables goods sector. Firms can hedge against these risks in several ways.

Firms with strong brand loyalty and differentiated products are less vulnerable to swings in consumer confidence. When consumer confidence declines, individuals tend to be more selective about their purchases, favoring trusted or indispensable brands. For example, Apple or luxury car manufacturers can rely on brand loyalty to sustain demand, even when general consumer confidence is low.

By offering products that appeal to a broad range of consumers, firms can hedge against declines in consumer confidence. When high-end consumers cut back on spending, demand from lower-income segments for more affordable products can help balance overall sales. For instance, a car manufacturer might hedge by offering both luxury and budget models, ensuring that it can cater to both premium and cost-conscious consumers during periods of low confidence.

During periods of low consumer confidence, firms can stimulate demand through targeted marketing and promotional strategies. Offering discounts, financing options, or loyalty rewards can encourage consumers to make purchases, even in uncertain times. For example, firms often introduce aggressive discounting strategies or "buy now, pay later" options to attract hesitant consumers during times of economic uncertainty.

Firms that operate in multiple regions or countries can hedge against domestic declines in consumer confidence by tapping into stronger or more stable international markets. Geographical diversification allows firms to mitigate the impact of localized economic downturns. For example, a global consumer electronics firm may experience reduced sales in the U.S. during a recession but find stronger demand in emerging markets where consumer confidence is higher.

The panel quantile Granger causality tests (PQGC) reported in Table 4 further bolster the findings of the PQR, confirming the robustness of the findings PQR estimation related to uncertainty indices. In each case the uncertainties variables Granger cause q as indicted by the Wald statistics that are significant at the 1 percent level. To further examine the robustness of our findings up to this point we estimate the panel vector autoregressive MIDAS model in the following section.

## SUMMARY AND CONCLUSIONS

This paper examines the relationship between macroeconomic variables—economic policy uncertainty (EPU), inflation risk, recessionary risks, and consumer confidence—and Tobin's Q for twelve U.S. consumer durable firms. Tobin's Q, a critical metric for firm valuation, incorporates both market value and replacement cost, making it a reliable indicator of a company's worth, particularly for firms with diverse capital structures.

The connection between economic uncertainty and firm value is well-grounded in economic theory. Early studies, such as Sandmo's seminal work in 1971, explored how uncertainty influences corporate behavior. Galbraith (1977) further emphasized the pivotal role of uncertainty in the financial and economic realms.

Building on this foundation, researchers like Flacco and Kroetch (1986), Fooladi and Kayhani (1991), and Adrangi and Raffiee (1999) have analyzed how firms navigate market risks. Their studies provide insights into how companies adjust production processes, pricing strategies, and profit-maximizing approaches in response to economic uncertainties.

In recent times, the research conducted by Baker et al. (2016) has been instrumental in quantifying and examining economic and economic policy uncertainties. Their efforts have involved the creation of indices, offering valuable perspectives into the ever-evolving landscape of economic uncertainty. These insights have proven invaluable for policymakers, economists, and investors, enhancing their comprehension of these vital aspects.

Firms in the durable consumer goods sector, which produce long-lasting products like automobiles, appliances, and furniture, are highly sensitive to macroeconomic conditions. Four key factors—economic policy uncertainty, inflationary uncertainty, recession risk, and consumer confidence—play a significant role in influencing the performance and decision-making of these firms.

Firms in the U.S. durable goods industry face significant risks from inflation, recessions, and shifts in consumer confidence, but can mitigate these through effective hedging strategies. These strategies include commodity hedging, cost management, revenue diversification, building liquidity, and marketing tactics to sustain demand, ensuring resilience during economic uncertainty.

Inflation increases input costs and price volatility. Durable goods producers, reliant on raw materials (e.g., metals, oil, agriculture), can hedge by using commodity futures contracts to lock in prices. Additionally, inflation-adjustment clauses in contracts allow firms to raise prices in response to rising costs, preserving margins. Diversifying suppliers or substituting more stable materials can also reduce inflation exposure. Firms with strong brand loyalty may pass higher costs onto consumers by raising prices, especially if demand is inelastic.

Firms in cyclical industries are vulnerable to recessions, which reduce consumer spending and increase credit costs. To hedge, firms should maintain strong liquidity by building cash reserves or securing lines of credit to weather downturns. A flexible cost structure, with variable costs like temporary workers or contract labor, allows quick scaling during reduced demand. To manage rising borrowing costs, firms should reduce high debt levels or hedge against interest rate hikes using swaps or fixed-rate loans.

Fluctuating consumer confidence directly affects demand for durable goods. Firms can hedge by fostering brand loyalty and offering differentiated products, making them less vulnerable to shifts in consumer sentiment. For example, brands like Apple maintain demand even during downturns. Offering a range of products, from premium to budget options, helps firms balance sales across different consumer segments. Marketing strategies such as discounts, financing options, or loyalty rewards can also stimulate demand in uncertain times. Additionally, firms operating internationally can mitigate domestic declines in consumer confidence by tapping into stronger international markets.

The findings of the PVM methodology, based on quarterly data from 1985 to 2022 for financial factors and Tobin's Q of these companies, as well as monthly data for the uncertainty variables, reveal that Q reacts negatively to economic policy uncertainty (EPU), recessionary risks, and inflationary expectations, while being positively associated with consumer confidence in this sample. The estimation of the PQR model corroborates these observations, particularly at the higher quantiles of the Q distribution. Granger causality tests further strengthen these findings.

The economic uncertainties triggered by the tariff policies introduced by the Trump administration in early 2025, along with the resulting shifts in both domestic and global economic policies, reinforce the findings of our study. Our research demonstrates that firm values—reflected in equity market capitalization and driven by expectations of future cash flows—are negatively impacted by heightened economic and policy uncertainty. These real-world developments serve as timely validation of our empirical results, highlighting the sensitivity of firm valuation to unpredictable policy environments.

Our results have several managerial implications for firms in the consumer goods sector. Specifically, we provide strategies that companies in this industry can implement to safeguard against shocks and uncertainties that threaten their financial stability and overall valuation. These strategies are designed to enhance resilience and ensure long-term growth in the face of volatile economic conditions. These proactive strategies enable durable goods firms to navigate economic volatility and continue delivering value to stakeholders.

**Declarations**
Authors declare no funding or conflict of interest
Data for the research are available upon request
**Acknowledgment**
We are grateful to anonymous reviewers for their valuable comments and suggestions. Remaining errors are the authors' responsibility.

Hendricks, W., & Koenker, R. (1992). Hierarchical spline models for conditional quantiles and the demand for electricity. *Journal of the American statistical Association*, *87*(417), 58-68.

Javier Fadul(2004). Business Ethics, Corporate Social Responsibility, and Firm Value in the Oil and Gas Industry. Doctoral Dissertation, Walden University, UMI Number: 3151491

Jin, Y., & Jorion, P. (2006). Firm value and hedging: Evidence from US oil and gas producers. *The journal of Finance*, *61*(2), 893-919.

Kato, K., Galvao Jr, A. F., & Montes-Rojas, G. V. (2012). Asymptotics for panel quantile regression models with individual effects. *Journal of Econometrics*, *170*(1), 76-91.

Koenker, R., & Bassett Jr, G. (1978). Regression quantiles. *Econometrica: journal of the Econometric Society*, 33-50.

Koenker, R. W., & d'Orey, V. (1987). Algorithm AS 229: Computing regression quantiles. *Applied statistics*, 383-393.

Koenker, R., & Machado, J. A. (1999). Goodness of fit and related inference processes for quantile regression. *Journal of the american statistical association*, *94*(448), 1296-1310.

Koenker, R. (2004). Quantile regression for longitudinal data. *Journal of multivariate analysis*, *91*(1), 74-89.

Koenker, R. (2005). Quantile regression [M]. *Econometric Society Monographs, Cambridge University Press, Cambridge*.

Koenker, R., & Ng, P. (2005). Inequality constrained quantile regression. *Sankhyā: The Indian Journal of Statistics*, 418-440.

Lamarche, C. (2010). Robust penalized quantile regression estimation for panel data. *Journal of Econometrics*, *157*(2), 396-408.

Lipsitz, S. F. (1997). Generalized Estimating Equations for Longitudinal Binary Data. *Journal of the American Statistical Association*, 998-1007.

Litterman, R. B. (1986). Forecasting with Bayesian vector autoregressions: Five years of experience. In V. (. Zarnowitz, *Business Cycles, Indicators, and Forecasting* (pp. 87–156). Chicago: University of Chicago Press.

Moon, H. R., & Weidner, M. (2015). Linear regression for panel with unknown number of factors as interactive fixed effects. *Econometrica*, *83*(4), 1543-1579.42

Table 1:  The list of the sample companies

| |
|---|
| BASSETT FURNITURE INDS |
| BUSH INDUSTRIES  -CL A |
| DEL LABORATORIES INC |
| EASTMAN KODAK CO |
| FIRST CITY INDUSTRIES INC |
| LA-Z-BOY INC |
| LEGGETT & PLATT INC |
| SMITH (A.O.) |
| IMAX Corp |
| WHIRLPOOL CORP |
| HARLEY-DAVIDSON INC |
| ETHAN ALLEN INTERIORS INC |

Table 2: Unit root tests and descriptive statistics  Q1/1985 to Q4/2022

Panel A: LLC unit Root test: Financial Variables

|     | cr | DA | OIAD | q | qr |
| --- | --- | --- | --- | --- | --- |
| LLC | -2.837 [a] | 2.323 | 1.677 | 0.48 | -2.576 [a] |

Panel B: Standard Unit root test: Uncertainty Variables  1985 to 2022

|     | EPU | Rec_Risk | INFEXP | Consconf |
| --- | --- | --- | --- | --- |
| ADF | -4.932 [a] | -4.347 [a] | -5.303 [a] | -3.195 [b] |
| PP  | -5.190 [a] | -4.400 [a] | -5.099 [a] | -2.900 [b] |

Notes:

Panel B: Descriptive statistics

|         | cr | DA | OIAD | q | qr |
| --- | --- | --- | --- | --- | --- |
| Average | 2.328 | 0.233 | 80.180  | 1.868   | 1.425 |
| Median  | 1.215 | 0.205 | 16.296  | 1.552   | 1.215 |
| Sd      | 0.833 | 0.195 | 143.110 | 2.363   | 0.833 |
| Max     | 6.619 | 1.302 | 914.000 | 80.467  | 6.619 |
| Min     | 0.079 | 0.000 | -301.000| -20.901 | 0.079 |

Notes:  Levine, Lin and chu t*( LLC ) unit root test assumes cross-sectional independence among variables of the panel data.  The null hypotheses for ADF and PP tests is that a series is nonstationary. Intercepts were included in the ADF and PP tests.  [a, b] represent significance at 1 and 5 percent levels.

Table 3: Panel Quantile Regression Estimation Results

|  | Quantiles | | |
|---|---|---|---|
| q | 25 | 50 | 75 |
| cr | -0.141 | -0.241 | -0.5960 |
|  | (0.0007) | (0.0003) | (0.0001) |
| qr | 0.256 | 0.2941 | 0.6449 |
|  | (0.0001) | (0.0003) | (0.0004) |
| da | 0.422 | -0.2925 | -1.2255 |
|  | (0.0001) | (0.0007) | (0.0002) |
| oiad | 0.001 | 0.0009 | 0.0020 |
|  | ($1.44*e-0.07$) | ($8.72*e-0.07$) | ($6.45*e-0.07$) |
| $epu_{t-1}$ | -0.0004 | -0.0004 | -0.0003 |
|  | ($4.08*e-0.07$) | ($1.44*e-0.07$) | ($5.04*e-0.07$) |
| $epu_{t-2}$ | 0.00009 | 0.00007 | -0.0005 |
|  | ($4.24*e-0.07$) | ($8.49*e-0.07$) | ($4.19*e-0.07$) |
| $epu_{t-3}$ | 0.0001 | 0.0003 | 0.0003 |
|  | ($4.98*e-0.07$) | ($1.54*e-0.06$) | ($9.33*e-0.07$) |
| $rec\_risk_{t-1}$ | 0.00009 | -0.00118 | -0.0026 |
|  | ($2.35*e-0.06$) | ($1.48*e-0.06$) | ($2.51*e-0.06$) |
| $rec\_risk_{t-2}$ | -0.0004 | 0.00127 | 0.0010 |
|  | ($2.40*e-0.06$) | ($6.62*e-0.06$) | ($4.30*e-0.06$) |
| $rec\_risk_{t-3}$ | -0.0011 | -0.00244 | -0.00254 |
|  | ($8.4*e-0.07$) | ($6.35*e-0.06$) | ($4.15*e-0.06$) |
| $infexp_{t-1}$ | -0.1154 | -0.2185 | -0.2641 |
|  | (0.00003) | (0.0003) | (0.0001) |
| $infexp_{t-2}$ | -0.0110 | -0.03596 | -0.1195 |
|  | (0.00017) | (0.00022) | (0.0002) |
| $infexp_{t-3}$ | -0.0946 | -0.03967 | -0.06498 |
|  | (0.00013) | (0.00006) | (0.00007) |
| $consconf_{t-1}$ | 0.0043 | -0.0036 | 0.000277 |
|  | ($2.32*e-0.06$) | (0.00011) | (0.00001) |
| $consconf_{t-2}$ | -0.0277 | 0.00013 | -0.00666 |
|  | (0.00001) | (0.000013) | ($4.64*e-0.06$) |
| $consconf_{t-3}$ | -0.0025 | 0.0025 | 0.00895 |
|  | ($6.63*e-0.06$) | ($6.98*e-0.06$) | (0.00002) |
| MCMC Diagnostics: Mean acceptance rate | | | |
|  | 0.002 | 0.004 | 0.003 |
| Obj. ft. | -1728.772 | -220.267 | -175.095 |

Notes: PQRs are estimated by Markov Chain Monte Carlo (MCMC) methodology. All estimated coefficients are statistically significant at 0.01 percent or less.

Table 4: Wald test of Granger Causality

| Null Hypothesis | Chi-sq |
|---|---|
| EPU Does not Granger Cause q | 4.9*e+0.07 [a] |
| Rec_risk Does not Granger Cause q | 1.4*e+0.07 [a] |
| infexp Does not Granger Cause q | 5.9*e+0.08 [a] |
| consconf Does not Granger Cause q | 3.2*e+10 [a] |

Notes: [a, b] represent significance at 1 and 5 percent levels. Significance of the Wald statistic leads to the rejection of the null hypothesis.

**Figure 1: Impulse response Analysis from the PVM model estimation**

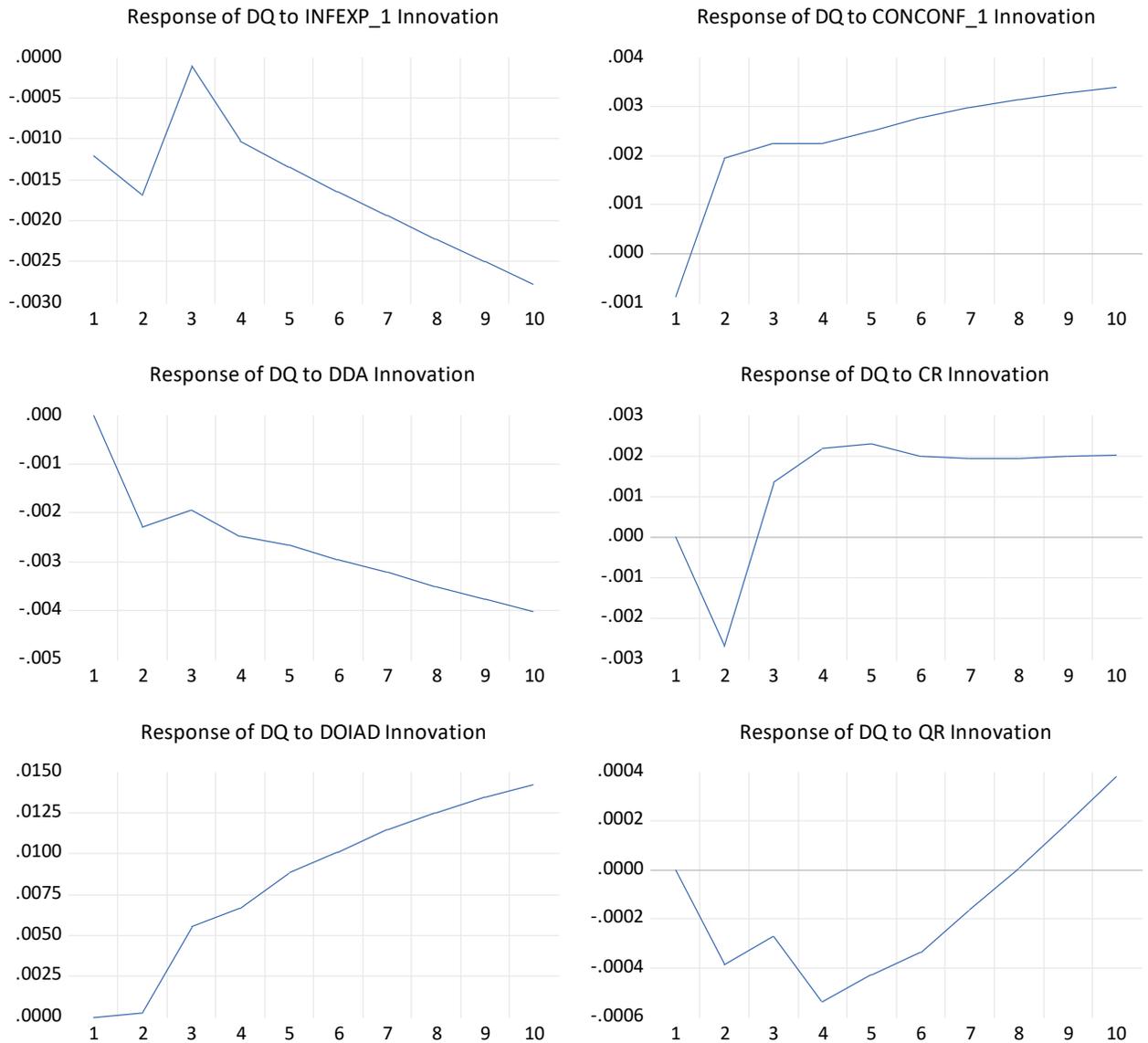

Notes: The PVRA-MIDAS model is estimated after necessary transformation of some model variables to ensure stationarity. The optimum lag length of the PVM is determined at 2 based on AIC, SC, FPE and HQ values.